\documentclass[%
 reprint, 
 superscriptaddress,
 amsmath,amssymb,
 aps,
floatfix,
longbibliography
]{revtex4-1}

\usepackage{graphicx}
\usepackage{bm}

\usepackage{xcolor}
\usepackage[colorlinks=true,citecolor=blue,linkcolor=magenta]{hyperref}

\newcommand{\la}{\langle}
\newcommand{\ra}{\rangle}
\newcommand{\rar}{\rightarrow}

\newcommand{\ben}{\begin{eqnarray}}
\newcommand{\een}{\end{eqnarray}}

\newcommand{\be}{\begin{equation}}
\newcommand{\ee}{\end{equation}}

\begin{document}

\title{Coherent reaction between molecular and atomic Bose-Einstein condensates: integrable model}
\author{Rajesh K. Malla}
\affiliation{Theoretical Division, and the Center for Nonlinear Studies, Los Alamos National Laboratory, Los Alamos, New Mexico 87545, USA}

\author{Vladimir Y. Chernyak}
\affiliation{Department of Chemistry, Wayne State University, 5101 Cass Ave, Detroit, Michigan 48202, USA}
\affiliation{Department of Mathematics, Wayne State University, 656 W. Kirby, Detroit, Michigan 48202, USA}
\author{Chen Sun}
\affiliation{School of Physics and Electronics, Hunan University, Changsha 410082, China}
\author{Nikolai A. Sinitsyn}
\affiliation{Theoretical Division, Los Alamos National Laboratory, Los Alamos, New Mexico 87545, USA
}
\email{nsinitsyn@lanl.gov}
\begin{abstract}

     We solve a model that describes a stimulated conversion between ultracold bosonic atoms and molecules. The reaction is triggered by a linearly time-dependent transition  throughout the Feshbach resonance. Our solution predicts a nonexponential dependence, with a dynamic phase transition, of the reaction efficiency on the transition rate. We find that the emerging phase can have a thermalized energy distribution with the temperature defined by the rate of the transition. This phase, however, has strong purely quantum correlations.
\end{abstract}

\maketitle

Recently,  a coherent conversion between  Cs atomic Bose-Einstein condensate (BEC) and the condensate of molecules Cs$_2$  was demonstrated
by changing  a magnetic field in time that pushed the ultracold atoms across the Feshbach resonance \cite{exp-21nat}. To suppress irreversible scatterings, the experimentalists in \cite{exp-21nat} confined their rotating BECs in a quasi-2D trap. In the  created  state, possibly up to 50\% of molecules formed a condensate, and in the reversed process, about 40\% of the molecules coherently dissociated into atoms.  Further improvements of  the conversion efficiency are expected. This will enable  {\it the chemistry of  coherent BECs} for  flexible engineering of macroscopic correlated quantum states, with applications in sensing and information-processing.

Every established branch of  physics has benefited from solving certain models without any approximations. For example, the solution for the hydrogen atom explains the  basic  observations in atomic physics, for which it serves as a starting point for numerous approximate techniques.

In this Letter, we  solve a model that can play a similar role in the chemistry of BECs.
 Its  Hamiltonian describes a stimulated conversion between bosonic atoms and molecules during a sweep of a linearly time-dependent magnetic field across a narrow Feshbach resonance. Without  approximations, the model captures the main features of the process: many-body interactions, energy dispersion of the atomic states, different initial populations of these states, and an arbitrary passage rate:
 \begin{eqnarray}
\label{H2}
\nonumber H(t)=-\beta t \hat{\Psi}^{\dagger} \hat{\Psi} +\sum_k \Big \{ \varepsilon^a_k   \hat{a}_k^{\dagger} \hat{a}_k+\varepsilon^b_k   \hat{b}_k^{\dagger} \hat{b}_k\\ +g\left(\hat{\Psi}^{\dagger}\hat{a}_k\hat{b}_k  +\hat{\Psi}\hat{a}^{\dagger}_k\hat{b}^{\dagger}_k \right)\Big \}.
\end{eqnarray}
Here, $\hat{a}_k$, $\hat{b}_k$ and $\hat{\Psi}$ are the boson annihilation operators;   $\varepsilon^{a,b}_k$ are the energies of the free atoms; $\beta$ is proportional to the ramp of the magnetic field that sweeps the system across the Feshbach resonance, and $g$ is the coupling for the  conversion of the atomic pairs into the molecules; $\hat{\Psi}$ is the molecular field, and $\hat{a}_k$ and $\hat{b}_k$ describe the modes of the ultracold atoms. They are generally different due to  momentum and spin conservation \cite{kayali}. However, we allow for  any reaction channel $k$ to identify $\hat{b}_k$ with $\hat{a}_k$. 
 
\begin{figure*}
\includegraphics[scale=0.15]{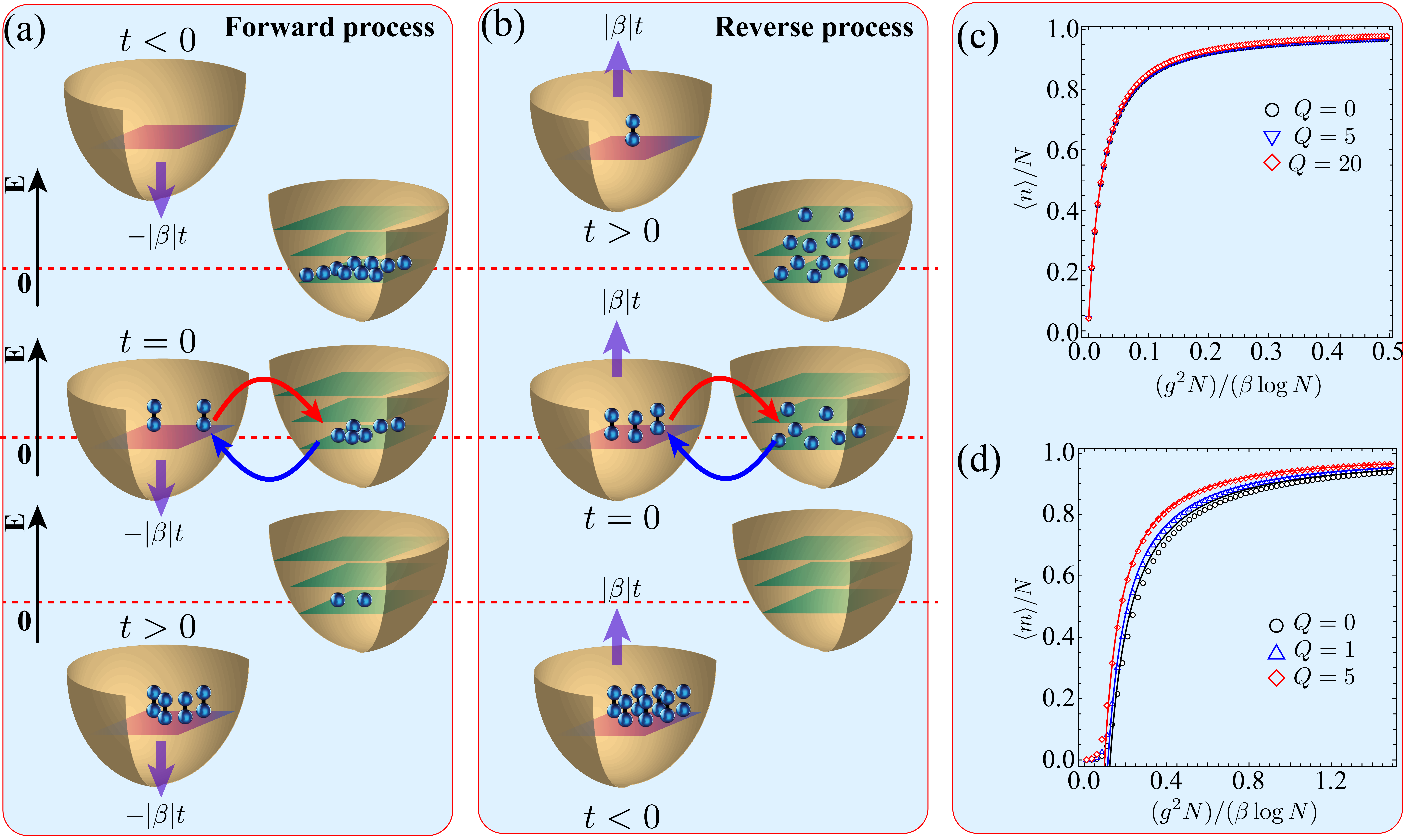}
\caption{The reaction between the atoms and the molecular condensate in (a) the forward and (b) the reverse process. The {\bf E}-axis is energy; violet arrows show the direction of the transition throughout the resonance. In the potential trap, molecular and atomic modes are shown as pink and green surfaces, respectively. (c) and (d) characterize  the average number of molecules  $\langle n\rangle$ and  of atomic pairs $\langle m\rangle=N-\langle n\rangle$ that are found for different values of the inverse sweep rate, $1/\beta$, at $g=1$, 
after the transitions starting from the ground state in the forward and the
backward processes in  (\ref{H11}), respectively. The discrete points are the exact predictions for the initial number of particles $N=10^4$ \cite{suppl}, and the
solid curves in (c) and (d) are the large-$N$ predictions of, respectively, Eq.~(\ref{mod-for}) and Eq.~(13) from \cite{suppl}. }
\label{Fig1}
\end{figure*}

 We assume that as $t\rightarrow -\infty$ the system is close to its ground state. Hence, (\ref{H2}) for $\beta>0$ describes the driven transition from atoms as $t\rightarrow -\infty$ into an initially empty  molecular mode [Fig.~\ref{Fig1}(a)]. We will call this a {\it forward process}.  The {\it reverse process}, of the molecular condensate dissociation in Fig.~\ref{Fig1}(b), corresponds to $\beta<0$, so that the ground state with a molecular condensate is at $t=-\infty$, and it is coherently converted into atomic pairs. Our goal is to find the final state as $t\rar +\infty$.

 The creation of ultracold molecules by sweeping the magnetic field across the resonance has been studied theoretically~\cite{yurovsky,altland-LZ,DTCM1,itin}, and  used experimentally~\cite{cond-exp1,cond-exp2,coldmol1,fesh-exp3,fesh-exp4} for two decades.  However, a  considerable fraction of the molecular condensate, until the experiment~\cite{exp-21nat}, was found only for the reactions of fermionic  atoms, which had fewer than bosonic possibilities to create detrimental excitations. Therefore, the purely bosonic reactions are yet  to be understood.

 The Hamiltonian $H(t)$ is explicitly time-dependent and has cubic interactions. There are no restrictions on all of its parameters, so (\ref{H2}) can describe both slow (small $\beta$) and fast (large $\beta$) transitions. The dependence on $\beta$ has been accessible both in \cite{exp-21nat} and previous experiments with fermion-molecular reactions. Hence, here we consider the questions: how the molecular-atomic conversion efficiency depends on $\beta$ and how by tuning $\beta$ one can control properties of the emerging state after the reaction?

To solve the model, we note that $Q_k=\hat{a}_k^{\dagger}\hat{a}_k-\hat{b}^{\dagger}_k\hat{b}_k$ commutes with $H(t)$, and  introduce operators
$K_k^+\equiv \hat{a}_k^{\dagger}\hat{b}_k^{\dagger}$ and $ K_k^-\equiv \hat{a}_k\hat{b}_k$, which satisfy 
\be
[\hat{n}_k,K^{\pm}_k]=\pm 2K_k^{\pm}, \quad [K_k^-,K_k^+]=(\hat{n}_k+1),
\label{com1}
\ee
where 
$\hat{n}_k\equiv \hat{a}^{\dagger}_k\hat{a}_k+\hat{b}^{\dagger}_k\hat{b}_k.
$
We also introduce parameters $\varepsilon_k=(\varepsilon_k^a+\varepsilon_k^b)$, and $\tau$ to rewrite $H(t)$ as 
\be
\label{h23}
H(t)=-\beta t \hat{\Psi}^{\dagger} \hat{\Psi} +\sum_k \big \{ \tau \varepsilon_k   \hat{n}_k/2+g\left[\hat{\Psi}^{\dagger}K^-_k  +\hat{\Psi}K^+_k \right]\big \}.
\ee
At $\tau=1$, we reproduce (\ref{H2}) up to $Q_k$-dependent terms that do not change the dynamics. Our main observation is that $H(t)$ in (\ref{h23}) commutes with 
\begin{eqnarray}
\label{h23-c}
\nonumber H'(t)&=&\sum_k \big \{ \varepsilon_k (t+\frac{\tau \varepsilon_k}{\beta})\frac{\hat{n}_k}{2}   +\frac{g \varepsilon_k}{\beta}\left[\hat{\Psi}^{\dagger}K^-_k  +\hat{\Psi}K^+_k \right] \big \} +
\\
&+&\frac{g^2}{\beta\tau}\sum_{i,j;\, i\ne j} \left(K_i^+K_j^- -(\hat{n}_i+1)(\hat{n}_j+1)/4\right),
\end{eqnarray}
and these  two operators satisfy the relation 
\be
\partial H/\partial \tau = \partial H'/\partial t.
\label{integ1}
\ee
According to \cite{commute,parallel-LZ}, this makes our model an integrable multistate Landau-Zener model, for which  any rescaling by  $\tau$  does not change the scattering probabilities between any eigenstates of $H(t)$. 

Hence, without affecting the final result, we can set $\tau \rightarrow \infty$, which renders all atomic states well-separated by times of their individual resonant interaction with the molecular mode. Near each such resonance, we can safely disregard all other reaction channels and then treat their effects sequentially in their chronological order. 

The Hamiltonian restricted  to a single reaction channel, $\hat{\Psi}\leftrightarrow \hat{a}\hat{b}$,  is given by
\begin{eqnarray}
\label{H11}
 H(t)=-\beta t \hat{\Psi}^{\dagger} \hat{\Psi} + g\left(\hat{\Psi}^{\dagger}\hat{a}\hat{b}  +\hat{\Psi}\hat{a}^{\dagger}\hat{b}^{\dagger} \right),
\end{eqnarray}
where we shifted the timescale to set the resonance at $t=0$.
The dynamics with (\ref{H11}) conserves two quantities:
\be
\label{cons1}
N=\hat{\Psi}^{\dagger}\hat{\Psi} + \hat{b}^{\dagger} \hat{b}, \quad 
Q=\hat{a}^{\dagger} \hat{a} -\hat{b}^{\dagger} \hat{b}.
\ee
This allows us to express the microstates only via the the number of molecules $n$:
$
|n\ra \equiv |n;N-n+Q, N-n\ra,$
where $N-n+Q$ and $N-n$ are the numbers of atoms in $a$ and $b$ modes, respectively. Note also that  at $Q=1$ (\ref{H11}) has the same matrix elements as the Hamiltonian with a single atomic mode: ${\hat a}\equiv {\hat b}$.

Our second observation is that  (\ref{H11}) can be mapped to the {\it  driven Tavis-Cummings model} (DTCM) \cite{DTCM,DTCM1}:
\be
H_{TC}=\beta t (\hat{\psi}^{\dagger} \hat{\psi}-N) +g(\hat{\psi}^{\dagger} \hat{S}^- +\hat{\psi} \hat{S}^+), 
\label{tc1}
\ee
where $\hat{\psi}$ is a boson annihilation operator, and $\hat{S}^{\pm}$ are raising/lowering operators of a spin with size $S=(N+Q)/2$.
The DTCM conserves 
$
N=\hat{\psi}^{\dagger} \hat{\psi}+(S+\hat{S}_z).
$
Hence, we can mark its states as $|n\ra$ where 
$
n=S+S_z,
$
Then,
\be
\la n+1|H_{TC}|n\ra =\sqrt{(N-n)(N+Q-n)(n+1)}.
\label{tc-me}
\ee
Comparing with (\ref{H11}), we find  
$$
\la n|H|n'\ra=\la n|H_{TC}|n'\ra, \quad \forall n,n'.
$$
 This map is not intuitive in the sense  that  the number of molecules $n$ in (\ref{H11}) is not the same as the number of bosons in the DTCM. Instead,  $\langle n|\hat{\psi}^{\dagger}\hat{\psi}|n\rangle=N-n$. Therefore, the dynamics that was called the forward process in DTCM~\cite{laser-LZ} is mapped to the reverse process in (\ref{H11}), and vice versa. Thus, we predict strongly different efficiencies in purely bosonic versus fermion-bosonic reactions. 

{\it Reaction efficiency}. The DTCM has been solved previously \cite{DTCM}. Its transition probabilities between any  microstates can be found in Eqs.~(23-24) in \cite{DTCM1}, where the limits of large $N$ are also described. Using our map to this model, we find for the forward sweep~\cite{suppl} that if all atoms are initially in the ground state with some number of pairs $N\gg1$ and $Q=O(1)$ then the probability distribution of finding $n$-molecules, for $\langle n \rangle \gg 1$, at the end is sharply peaked near the average value
\begin{equation}
\langle n \rangle\equiv\langle \hat{\Psi}^{\dagger} \hat{\Psi}\rangle_{t\rightarrow \infty}=N+\frac{\log{\left(2-x^{N+Q}\right)}}{\log{x}},\quad  x=e^{-\frac{2\pi g^2}{{\beta}}}.
   \label{mod-for}
\end{equation}

For the reverse sweep, starting with $N$ molecules and no atoms, after the passage through one resonance, the probability to produce $m$ atomic pairs is  given by \cite{suppl}
\begin{equation}
    P_m=x^{N-m}(x^{N-m+1},x)_m, \quad x=e^{-2\pi g^2/\beta},
    \label{one-res}
\end{equation}
where 
\begin{equation}
    (a,x)_m\equiv \prod_{k=0}^{m-1}(1-ax^k) = (1-a)(1-ax)\cdots(1-ax^{m-1})
    \label{qPotch-def}
\end{equation}
is the q-Pochhammer symbol.
According to~\cite{laser-LZ}, this distribution predicts a dynamic phase transition. Namely, if the number of molecules $N$ is initially macroscopically large, then the fraction of molecules converted to atoms, 
$
\langle m \rangle /N =1-\langle n \rangle/N,
$
behaves discontinuously as a function of, $g^2/\beta$. Let 
\be
f=\frac{2\pi g^2}{\beta}\frac{N}{{\rm log_e}N},
\label{f-def}
\ee
then the distribution (\ref{one-res}) has the property
\begin{eqnarray}
\nonumber \langle m \rangle/N &=&0 \quad {\rm for } \quad f<1,\quad N\rightarrow \infty, \\
\label{nbar-1}
\langle m \rangle/N &=&\frac{f-1}{ f} \quad {\rm for } \quad f\ge 1,\quad N\rightarrow \infty.
\end{eqnarray}

In Figs.~\ref{Fig1}(c,d) we confirm the predictions (\ref{mod-for}),~(\ref{nbar-1}) using numerically exact transition probabilities \cite{suppl} for $N=10^4$, and also verify robustness of such predictions against an initial asymmetry in atomic population due to nonzero $Q$. Here we  note similarity of Fig.~\ref{Fig1}(d) with the experimentally obtained Fig.4(a) in \cite{exp-21nat}. Both figures describe the number of  produced atoms from molecules: in our case as a function of $1/\beta$ but as a function of time of the resonant interaction in \cite{exp-21nat}. We attribute this to the fact that $g/\beta$ in our model characterizes the effective time of the transition through the resonance, so both figures describe essentially the same physics: it takes initially certain critical time for the process to evolve without visible effects, after which a macroscopic number of molecules dissociates quickly due to superradiance \cite{laser-LZ}, which can occur before entering stationary equilibrium phases \cite{phase-transition-1,phase-transition-2,angela1,angela2,angela3}.

{\it Coherent thermalization}. Consider now  the effect of the dispersion $\varepsilon_k$ of atomic modes in the general model (\ref{H2}). For the forward-sweep, the integrability means that only higher energy atomic pairs can influence the lower energy ones. Hence, if the initial state is the atomic ground state then the higher energy states remain empty at the end, and Eq.~(\ref{mod-for}) applies to this case as well.

For the reverse process, if the initial state has $N$ molecules, then Eq.~(\ref{one-res}) applies to the first encountered resonance. It also applies to the following resonances but the number of entering molecules must be reduced by the amount that has already dissociated. If there are many resonances, all molecules will dissociate. 

The final multi-mode population distribution  has a very simple structure. To show this, 
 we write the joint transition probability to produce $m_1$ atomic pairs in the first and $m_2$ pairs in the second resonance:
\begin{equation}
    P_{m_1,m_2}=P_{m_1} x^{N-{m_1}-m_2}(x^{N-m_1-m_2+1},x)_{m_2},
    \label{pnm-1}
\end{equation}
and compare two probabilities of the populations that are different by  moving one atomic pair from the lower energy mode to the higher energy one. Taking the ratio of such probabilities, we find  
\begin{equation}
  \frac{P_{m_1-1,m_2+1}}{P_{m_1,m_2}}=e^{-2\pi g^2/\beta},
  \label{detailedb}
\end{equation}
which does not depend on $m_1$ and $m_2$. The same is true for any pair of the nearest energy atomic modes. This means that all probabilities satisfy the detailed balance conditions that are found in the  Gibbs distribution:
\begin{equation}
P_{\{m_k\}} = \frac{1}{Z}e^{-\frac{1}{k_BT}\sum_{k=1}^{\infty}km_k}\delta\left(N-\sum_{k}m_k\right),    
    \label{P-tot}
\end{equation}
where $\{m_k\}$ is the vector of the final atomic $a_kb_k$-mode populations, assuming that the mode indices grow according to the ordering of  energies $\varepsilon_k$ in (\ref{h23}). 
Here, $Z$ is a normalization factor,   the delta function  follows from the particle conservation, and
\begin{equation}
    k_BT\equiv \beta/(2\pi g^2). 
    \label{kt1}
\end{equation}
The experiment \cite{exp-21nat}, indeed, found that after the reverse sweep the atomic energy population was thermalized at  elevated temperatures. Physically, however, (\ref{P-tot}) would be thermal only for a linear energy dispersion, $\varepsilon_k \sim k$. We attribute the agreement with~\cite{exp-21nat} to the fact that in its 2D trap geometry  the atomic energy dispersion is expected to be linear because 
$A\iint_0^\infty \frac{dk_xdk_y}{(2\pi)^2}\, \delta (E-k^2/(2m_a)) =A\int_0^\infty \frac{kdk}{(2\pi)}\delta (E-k^2/(2m_a))=Am_a/(2\pi),
$
where $m_a$ is the atomic mass, and $A$ is the area of the trap. Hence, we predict that the physical temperature is $k_BT =A\beta m_a/ (2\pi g)^2$. The physical coupling $g$ decays with $A$, so only the linear dependence on $\beta$ is our testable prediction. 
The dynamic phase transition in Fig.~\ref{Fig1}(d)  now has a new interpretation: the final atomic distribution coincides with the equilibrium one for free bosons, which form a BEC below a critical temperature.

\begin{figure}
\includegraphics[scale=0.37]{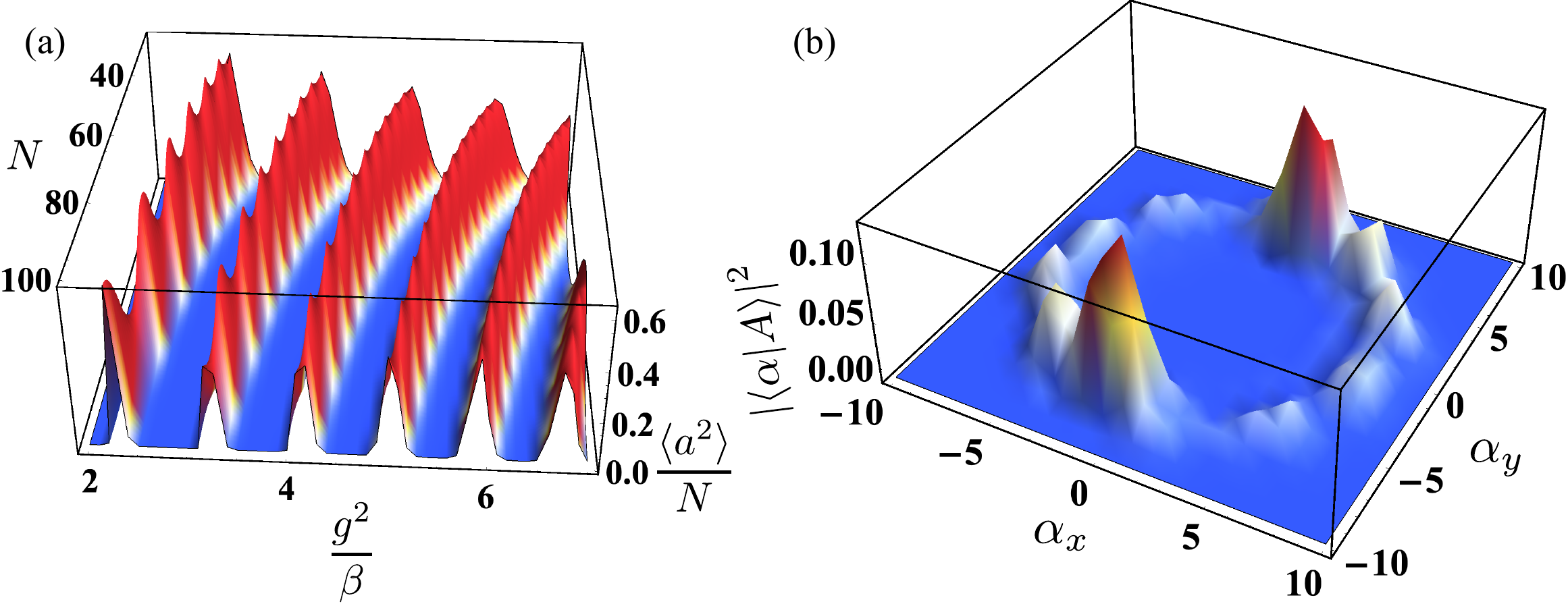}
\caption{(a) The ratio $|\langle A|\hat{a}^2|A\rangle|/N$ oscillates for changing $\beta$ and $N$. (b) 
The overlap of $|A\rangle$ with Glauber states  $|\alpha \rangle$, $\alpha=\alpha_{x}+i\alpha_{y}$, reveals two peaks. Here, the parameters for $|A\rangle$ are: $\phi_{M}=0$, $|\alpha_m|=g^2/\beta=5$. }
\label{fig-coh}
\end{figure}
{\it Phase coherence}. The Gibbs distribution (\ref{P-tot}) does not mean that the system after the reaction is truly thermalized because it is still described  by a coherent state vector. This can be revealed by measuring the relative phases of the final states, which can also be found exactly. In  \cite{suppl}, we show how they can be derived but we postpone the detailed analysis  to a follow-up work, and consider here only the adiabatic limit.  Although it is experimentally unreachable, our goal here is to show that the coherent behavior in the model is, indeed, rich.

Let the molecular condensate be in a coherent state: 
\begin{equation}
 |\alpha_m \rangle =e^{-|\alpha_m|^2/2}\sum_{n=0}^{\infty} \frac{\alpha_m^n}{\sqrt{n!}}|n\rangle.   
    \label{coh1}
\end{equation}
 The condensate phase $\phi_M$ is then defined  from $\langle \alpha_m| \hat{\Psi} |\alpha_m\rangle \sim \sqrt{N} e^{i {\rm arg} (\alpha_m)}$, where $N=|\alpha|^2$, so $\phi_M={\rm arg} (\alpha_m)$.
Consider the adiabatic dissociation of this condensate into a single atomic mode: $H=-\beta t\hat{\Psi}^{\dagger}\hat{\Psi} + g[\hat{\Psi}^{\dagger}\hat{a}^2 +\hat{\Psi}(\hat{a}^{\dagger})^2 ]$. Each $|n\rangle$ in (\ref{coh1}) is then converted to $e^{i\phi_n}|2n\rangle_a$, where $a$ marks the atomic states. In \cite{suppl} we derive an exact formula for the scattering phase of the complete dissociation amplitude:
\begin{equation}
    \phi_n=3n\pi/4-\sum_{k=1}^{n}arg\Gamma[i(g^2/\beta) (k+1)].
    \label{phase}
\end{equation}
For $g^2/\beta \gg 1$ it is simplified because, for $s\gg 1$, $\pi/4+arg\Gamma[is]\approx s({\rm log}_e s -1)$.
As $t\rightarrow +\infty$, the  atomic state becomes
\begin{equation}
    |A\rangle = e^{-|\alpha_m|^2/2}\sum_{n=0}^{\infty} \frac{(e^{3i\pi/4}\alpha_m)^n}{\sqrt{n!}}e^{i\phi_n}|2n\rangle_a. 
    \label{psif}
\end{equation}
This state contains information about the original molecular state. Thus, imagine that the molecular condensate  has a vortex, such that $\phi_M$ winds by $2\pi$ around some spatial point. This  topological property is preserved after the reaction because the circulation
\begin{equation}
-i\int_0^{2\pi} \langle A|\frac{d}{d\phi_M}|A\rangle \, d\phi_M = 2\pi N
\label{invt}
\end{equation}
 remains the same  as it would be for the initial state $|\alpha_m\rangle$. Hence, the stimulated reaction should preserve vortex-like spatial distributions of particles and currents. 

The state $|A\rangle$ is far from a coherent state because $\langle A| \hat{a} |A\rangle =0$. To test whether it can be close to a squeezed state, in Fig.~\ref{fig-coh}(a) we plot a numerically calculated ratio $|\langle A|\hat{a}^2 |A\rangle|/N$ as a function of $g^2/\beta$. Its values close to $1$ at large $N$ indicate the emergence of the squeezed state. We find that this ratio is generally small except at narrow resonant values of width $\sim 1/\sqrt{N}$ near $g^2{\rm log}_e(g^2N/
\beta)/\beta =2\pi n$, $n=1,2\ldots$, where it exceeds $0.5$.  Figure~\ref{fig-coh}(b) shows the overlap of a typical $|A\rangle$ with coherent states $|\alpha\rangle$, revealing  two peaks with opposite signs of the corresponding $\alpha$; this makes $|A\rangle$ akin to a macroscopic cat-state. As  $\phi_M$ changes from $0$ to $2\pi$, the axis connecting the two peaks rotates so that the peaks exchange their positions at the end. This is a topological consequence of the geometric phase (\ref{invt}).

In conclusion, our  solution provides a nonperturbative description of the reaction efficiency and quantum correlations for stimulated BEC reactions. It predicts a dynamic phase transition and  thermalization of the atomic energy distribution, which were observed experimentally in \cite{exp-21nat}. The emerging state after the reaction has considerable nonclassical correlations with properties that should be of interest for quantum sensing and topological quantum computing.


\section*{Acknowledgements}
The authors thank Avadh Saxena and Junyu Lin for useful discussions.
This work was  supported (R.K.M, V.Y.C, N.A.S.) by the U.S. Department of Energy, Office of Science, Basic Energy Sciences, Materials Sciences and Engineering Division, Condensed Matter Theory Program. The completion of this work was also supported (V.Y.C and N.A.S.) by the U.S. Department of Energy, Office of Science, Basic Energy Sciences, under Award Number DE-SC0022134. R.K.M. was partly supported by LANL  Center for Nonlinear Studies. C.S. contributed to the discussion that lead to the idea of this article. He was supported by NSFC (No. 12105094) and by the Fundamental Research Funds for the Central Universities from China.

\bibliography{ref}

\begin{thebibliography}{23}%
\makeatletter
\providecommand \@ifxundefined [1]{%
 \@ifx{#1\undefined}
}%
\providecommand \@ifnum [1]{%
 \ifnum #1\expandafter \@firstoftwo
 \else \expandafter \@secondoftwo
 \fi
}%
\providecommand \@ifx [1]{%
 \ifx #1\expandafter \@firstoftwo
 \else \expandafter \@secondoftwo
 \fi
}%
\providecommand \natexlab [1]{#1}%
\providecommand \enquote  [1]{``#1''}%
\providecommand \bibnamefont  [1]{#1}%
\providecommand \bibfnamefont [1]{#1}%
\providecommand \citenamefont [1]{#1}%
\providecommand \href@noop [0]{\@secondoftwo}%
\providecommand \href [0]{\begingroup \@sanitize@url \@href}%
\providecommand \@href[1]{\@@startlink{#1}\@@href}%
\providecommand \@@href[1]{\endgroup#1\@@endlink}%
\providecommand \@sanitize@url [0]{\catcode `\\12\catcode `\$12\catcode
  `\&12\catcode `\#12\catcode `\^12\catcode `\_12\catcode `\%12\relax}%
\providecommand \@@startlink[1]{}%
\providecommand \@@endlink[0]{}%
\providecommand \url  [0]{\begingroup\@sanitize@url \@url }%
\providecommand \@url [1]{\endgroup\@href {#1}{\urlprefix }}%
\providecommand \urlprefix  [0]{URL }%
\providecommand \Eprint [0]{\href }%
\providecommand \doibase [0]{http://dx.doi.org/}%
\providecommand \selectlanguage [0]{\@gobble}%
\providecommand \bibinfo  [0]{\@secondoftwo}%
\providecommand \bibfield  [0]{\@secondoftwo}%
\providecommand \translation [1]{[#1]}%
\providecommand \BibitemOpen [0]{}%
\providecommand \bibitemStop [0]{}%
\providecommand \bibitemNoStop [0]{.\EOS\space}%
\providecommand \EOS [0]{\spacefactor3000\relax}%
\providecommand \BibitemShut  [1]{\csname bibitem#1\endcsname}%
\let\auto@bib@innerbib\@empty
\bibitem [{\citenamefont {Zhang}\ \emph {et~al.}(2021)\citenamefont {Zhang},
  \citenamefont {Chen}, \citenamefont {Yao},\ and\ \citenamefont
  {Chin}}]{exp-21nat}%
  \BibitemOpen
  \bibfield  {author} {\bibinfo {author} {\bibfnamefont {Zhendong}\
  \bibnamefont {Zhang}}, \bibinfo {author} {\bibfnamefont {Liangchao}\
  \bibnamefont {Chen}}, \bibinfo {author} {\bibfnamefont {Kai-Xuan}\
  \bibnamefont {Yao}}, \ and\ \bibinfo {author} {\bibfnamefont {Cheng}\
  \bibnamefont {Chin}},\ }\bibfield  {title} {\enquote {\bibinfo {title}
  {Transition from an atomic to a molecular {Bose-Einstein} condensate},}\
  }\href {\doibase 10.1038/s41586-021-03443-0} {\bibfield  {journal} {\bibinfo
  {journal} {Nature}\ }\textbf {\bibinfo {volume} {592}},\ \bibinfo {pages}
  {708--711} (\bibinfo {year} {2021})}\BibitemShut {NoStop}%
\bibitem [{\citenamefont {Kayali}\ and\ \citenamefont
  {Sinitsyn}(2003)}]{kayali}%
  \BibitemOpen
  \bibfield  {author} {\bibinfo {author} {\bibfnamefont {M.~A.}\ \bibnamefont
  {Kayali}}\ and\ \bibinfo {author} {\bibfnamefont {N.~A.}\ \bibnamefont
  {Sinitsyn}},\ }\bibfield  {title} {\enquote {\bibinfo {title} {Formation of a
  two-component {Bose} condensate during the chemical-potential curve
  crossing},}\ }\href {\doibase 10.1103/PhysRevA.67.045603} {\bibfield
  {journal} {\bibinfo  {journal} {Phys. Rev. A}\ }\textbf {\bibinfo {volume}
  {67}},\ \bibinfo {pages} {045603} (\bibinfo {year} {2003})}\BibitemShut
  {NoStop}%
\bibitem [{\citenamefont {Yurovsky}\ \emph {et~al.}(2002)\citenamefont
  {Yurovsky}, \citenamefont {Ben-Reuven},\ and\ \citenamefont
  {Julienne}}]{yurovsky}%
  \BibitemOpen
  \bibfield  {author} {\bibinfo {author} {\bibfnamefont {V.~A.}\ \bibnamefont
  {Yurovsky}}, \bibinfo {author} {\bibfnamefont {A.}~\bibnamefont
  {Ben-Reuven}}, \ and\ \bibinfo {author} {\bibfnamefont {P.~S.}\ \bibnamefont
  {Julienne}},\ }\bibfield  {title} {\enquote {\bibinfo {title} {Quantum
  effects on curve crossing in a {Bose-Einstein} condensate},}\ }\href
  {\doibase 10.1103/PhysRevA.65.043607} {\bibfield  {journal} {\bibinfo
  {journal} {Phys. Rev. A}\ }\textbf {\bibinfo {volume} {65}},\ \bibinfo
  {pages} {043607} (\bibinfo {year} {2002})}\BibitemShut {NoStop}%
\bibitem [{\citenamefont {Altland}\ \emph {et~al.}(2009)\citenamefont
  {Altland}, \citenamefont {Gurarie}, \citenamefont {Kriecherbauer},\ and\
  \citenamefont {Polkovnikov}}]{altland-LZ}%
  \BibitemOpen
  \bibfield  {author} {\bibinfo {author} {\bibfnamefont {Alexander}\
  \bibnamefont {Altland}}, \bibinfo {author} {\bibfnamefont {V.}~\bibnamefont
  {Gurarie}}, \bibinfo {author} {\bibfnamefont {T.}~\bibnamefont
  {Kriecherbauer}}, \ and\ \bibinfo {author} {\bibfnamefont {A.}~\bibnamefont
  {Polkovnikov}},\ }\bibfield  {title} {\enquote {\bibinfo {title}
  {Nonadiabaticity and large fluctuations in a many-particle {Landau-Zener}
  problem},}\ }\href {\doibase 10.1103/PhysRevA.79.042703} {\bibfield
  {journal} {\bibinfo  {journal} {Phys. Rev. A}\ }\textbf {\bibinfo {volume}
  {79}},\ \bibinfo {pages} {042703} (\bibinfo {year} {2009})}\BibitemShut
  {NoStop}%
\bibitem [{\citenamefont {Sun}\ and\ \citenamefont {Sinitsyn}(2016)}]{DTCM1}%
  \BibitemOpen
  \bibfield  {author} {\bibinfo {author} {\bibfnamefont {Chen}\ \bibnamefont
  {Sun}}\ and\ \bibinfo {author} {\bibfnamefont {Nikolai~A.}\ \bibnamefont
  {Sinitsyn}},\ }\bibfield  {title} {\enquote {\bibinfo {title} {{Landau-Zener}
  extension of the {Tavis-Cummings} model: Structure of the solution},}\ }\href
  {\doibase 10.1103/PhysRevA.94.033808} {\bibfield  {journal} {\bibinfo
  {journal} {Phys. Rev. A}\ }\textbf {\bibinfo {volume} {94}},\ \bibinfo
  {pages} {033808} (\bibinfo {year} {2016})}\BibitemShut {NoStop}%
\bibitem [{\citenamefont {Itin}\ and\ \citenamefont {T\"orm\"a}(2009)}]{itin}%
  \BibitemOpen
  \bibfield  {author} {\bibinfo {author} {\bibfnamefont {A.~P.}\ \bibnamefont
  {Itin}}\ and\ \bibinfo {author} {\bibfnamefont {P.}~\bibnamefont
  {T\"orm\"a}},\ }\bibfield  {title} {\enquote {\bibinfo {title} {Dynamics of a
  many-particle {Landau-Zener} model: Inverse sweep},}\ }\href {\doibase
  10.1103/PhysRevA.79.055602} {\bibfield  {journal} {\bibinfo  {journal} {Phys.
  Rev. A}\ }\textbf {\bibinfo {volume} {79}},\ \bibinfo {pages} {055602}
  (\bibinfo {year} {2009})}\BibitemShut {NoStop}%
\bibitem [{\citenamefont {Herbig}\ \emph {et~al.}(2003)\citenamefont {Herbig},
  \citenamefont {Kraemer}, \citenamefont {Mark}, \citenamefont {Weber},
  \citenamefont {Chin}, \citenamefont {N{\"a}gerl},\ and\ \citenamefont
  {Grimm}}]{cond-exp1}%
  \BibitemOpen
  \bibfield  {author} {\bibinfo {author} {\bibfnamefont {Jens}\ \bibnamefont
  {Herbig}}, \bibinfo {author} {\bibfnamefont {Tobias}\ \bibnamefont
  {Kraemer}}, \bibinfo {author} {\bibfnamefont {Michael}\ \bibnamefont {Mark}},
  \bibinfo {author} {\bibfnamefont {Tino}\ \bibnamefont {Weber}}, \bibinfo
  {author} {\bibfnamefont {Cheng}\ \bibnamefont {Chin}}, \bibinfo {author}
  {\bibfnamefont {Hanns-Christoph}\ \bibnamefont {N{\"a}gerl}}, \ and\ \bibinfo
  {author} {\bibfnamefont {Rudolf}\ \bibnamefont {Grimm}},\ }\bibfield  {title}
  {\enquote {\bibinfo {title} {Preparation of a pure molecular quantum gas},}\
  }\href {\doibase 10.1126/science.1088876} {\bibfield  {journal} {\bibinfo
  {journal} {Science}\ }\textbf {\bibinfo {volume} {301}},\ \bibinfo {pages}
  {1510--1513} (\bibinfo {year} {2003})}\BibitemShut {NoStop}%
\bibitem [{\citenamefont {Xu}\ \emph {et~al.}(2003)\citenamefont {Xu},
  \citenamefont {Mukaiyama}, \citenamefont {Abo-Shaeer}, \citenamefont {Chin},
  \citenamefont {Miller},\ and\ \citenamefont {Ketterle}}]{cond-exp2}%
  \BibitemOpen
  \bibfield  {author} {\bibinfo {author} {\bibfnamefont {K.}~\bibnamefont
  {Xu}}, \bibinfo {author} {\bibfnamefont {T.}~\bibnamefont {Mukaiyama}},
  \bibinfo {author} {\bibfnamefont {J.~R.}\ \bibnamefont {Abo-Shaeer}},
  \bibinfo {author} {\bibfnamefont {J.~K.}\ \bibnamefont {Chin}}, \bibinfo
  {author} {\bibfnamefont {D.~E.}\ \bibnamefont {Miller}}, \ and\ \bibinfo
  {author} {\bibfnamefont {W.}~\bibnamefont {Ketterle}},\ }\bibfield  {title}
  {\enquote {\bibinfo {title} {Formation of quantum-degenerate sodium
  molecules},}\ }\href {\doibase 10.1103/PhysRevLett.91.210402} {\bibfield
  {journal} {\bibinfo  {journal} {Phys. Rev. Lett.}\ }\textbf {\bibinfo
  {volume} {91}},\ \bibinfo {pages} {210402} (\bibinfo {year}
  {2003})}\BibitemShut {NoStop}%
\bibitem [{\citenamefont {Bohn}\ \emph {et~al.}(2017)\citenamefont {Bohn},
  \citenamefont {Rey},\ and\ \citenamefont {Ye}}]{coldmol1}%
  \BibitemOpen
  \bibfield  {author} {\bibinfo {author} {\bibfnamefont {John~L.}\ \bibnamefont
  {Bohn}}, \bibinfo {author} {\bibfnamefont {Ana~Maria}\ \bibnamefont {Rey}}, \
  and\ \bibinfo {author} {\bibfnamefont {Jun}\ \bibnamefont {Ye}},\ }\bibfield
  {title} {\enquote {\bibinfo {title} {Cold molecules: Progress in quantum
  engineering of chemistry and quantum matter},}\ }\href {\doibase
  10.1126/science.aam6299} {\bibfield  {journal} {\bibinfo  {journal}
  {Science}\ }\textbf {\bibinfo {volume} {357}},\ \bibinfo {pages} {1002--1010}
  (\bibinfo {year} {2017})}\BibitemShut {NoStop}%
\bibitem [{\citenamefont {Matyja\ifmmode~\acute{s}\else \'{s}\fi{}kiewicz}\
  \emph {et~al.}(2008)\citenamefont {Matyja\ifmmode~\acute{s}\else
  \'{s}\fi{}kiewicz}, \citenamefont {Szyma\ifmmode~\acute{n}\else
  \'{n}\fi{}ska},\ and\ \citenamefont {G\'oral}}]{fesh-exp3}%
  \BibitemOpen
  \bibfield  {author} {\bibinfo {author} {\bibfnamefont {S.}~\bibnamefont
  {Matyja\ifmmode~\acute{s}\else \'{s}\fi{}kiewicz}}, \bibinfo {author}
  {\bibfnamefont {M.~H.}\ \bibnamefont {Szyma\ifmmode~\acute{n}\else
  \'{n}\fi{}ska}}, \ and\ \bibinfo {author} {\bibfnamefont {K.}~\bibnamefont
  {G\'oral}},\ }\bibfield  {title} {\enquote {\bibinfo {title} {Probing
  fermionic condensates by fast-sweep projection onto {Feshbach} molecules},}\
  }\href {\doibase 10.1103/PhysRevLett.101.150410} {\bibfield  {journal}
  {\bibinfo  {journal} {Phys. Rev. Lett.}\ }\textbf {\bibinfo {volume} {101}},\
  \bibinfo {pages} {150410} (\bibinfo {year} {2008})}\BibitemShut {NoStop}%
\bibitem [{\citenamefont {Duda}\ \emph {et~al.}(2021)\citenamefont {Duda},
  \citenamefont {Chen}, \citenamefont {Schindewolf}, \citenamefont {Bause},
  \citenamefont {Milczewski}, \citenamefont {Schmidt},\ and\ \citenamefont
  {Luo}}]{fesh-exp4}%
  \BibitemOpen
  \bibfield  {author} {\bibinfo {author} {\bibfnamefont {M.}~\bibnamefont
  {Duda}}, \bibinfo {author} {\bibfnamefont {X-Y.}\ \bibnamefont {Chen}},
  \bibinfo {author} {\bibfnamefont {A.}~\bibnamefont {Schindewolf}}, \bibinfo
  {author} {\bibfnamefont {R.}~\bibnamefont {Bause}}, \bibinfo {author}
  {\bibfnamefont {J.}~\bibnamefont {Milczewski}}, \bibinfo {author}
  {\bibfnamefont {I.}~\bibnamefont {Schmidt}, \bibfnamefont {R.~Bloch}}, \ and\
  \bibinfo {author} {\bibfnamefont {X-Y.}\ \bibnamefont {Luo}},\ }\bibfield
  {title} {\enquote {\bibinfo {title} {Transition from a polaronic condensate
  to a degenerate {Fermi} gas of heteronuclear molecules},}\ }\href@noop {}
  {\bibfield  {journal} {\bibinfo  {journal} {Preprit arXiv:2111.04301}\ }
  (\bibinfo {year} {2021})}\BibitemShut {NoStop}%
\bibitem [{sup()}]{suppl}%
  \BibitemOpen
  \bibfield  {title} {\enquote {\bibinfo {title} {Supplementary material, which
  additionally cites refs.~\cite{sinitsyn-chain,vitanov-parallel}, has four
  sections with (i) comments on the strongly nonadiabatic regime; (ii)
  derivation of the scattering phase for the complete molecular-atomic
  conversion amplitude; (iii) derivation of the continuous limit for the
  reaction rates at finite {$Q$}, and (iv) numerical checks for the predicted
  transition probabilities},}\ }\href@noop {} {\ }\BibitemShut {NoStop}%
\bibitem [{\citenamefont {Sinitsyn}\ \emph {et~al.}(2018)\citenamefont
  {Sinitsyn}, \citenamefont {Yuzbashyan}, \citenamefont {Chernyak},
  \citenamefont {Patra},\ and\ \citenamefont {Sun}}]{commute}%
  \BibitemOpen
  \bibfield  {author} {\bibinfo {author} {\bibfnamefont {Nikolai~A.}\
  \bibnamefont {Sinitsyn}}, \bibinfo {author} {\bibfnamefont {Emil~A.}\
  \bibnamefont {Yuzbashyan}}, \bibinfo {author} {\bibfnamefont {Vladimir~Y.}\
  \bibnamefont {Chernyak}}, \bibinfo {author} {\bibfnamefont {Aniket}\
  \bibnamefont {Patra}}, \ and\ \bibinfo {author} {\bibfnamefont {Chen}\
  \bibnamefont {Sun}},\ }\bibfield  {title} {\enquote {\bibinfo {title}
  {Integrable time-dependent quantum {Hamiltonians}},}\ }\href {\doibase
  10.1103/PhysRevLett.120.190402} {\bibfield  {journal} {\bibinfo  {journal}
  {Phys. Rev. Lett.}\ }\textbf {\bibinfo {volume} {120}},\ \bibinfo {pages}
  {190402} (\bibinfo {year} {2018})}\BibitemShut {NoStop}%
\bibitem [{\citenamefont {Chernyak}\ \emph {et~al.}(2020)\citenamefont
  {Chernyak}, \citenamefont {Li}, \citenamefont {Sun},\ and\ \citenamefont
  {Sinitsyn}}]{parallel-LZ}%
  \BibitemOpen
  \bibfield  {author} {\bibinfo {author} {\bibfnamefont {Vladimir~Y}\
  \bibnamefont {Chernyak}}, \bibinfo {author} {\bibfnamefont {Fuxiang}\
  \bibnamefont {Li}}, \bibinfo {author} {\bibfnamefont {Chen}\ \bibnamefont
  {Sun}}, \ and\ \bibinfo {author} {\bibfnamefont {Nikolai~A}\ \bibnamefont
  {Sinitsyn}},\ }\bibfield  {title} {\enquote {\bibinfo {title} {Integrable
  multistate {{Landau{\textendash}Zener} models} with parallel energy
  levels},}\ }\href {\doibase 10.1088/1751-8121/ab9464} {\bibfield  {journal}
  {\bibinfo  {journal} {Journal of Physics A: Mathematical and Theoretical}\
  }\textbf {\bibinfo {volume} {53}},\ \bibinfo {pages} {295201} (\bibinfo
  {year} {2020})}\BibitemShut {NoStop}%
\bibitem [{\citenamefont {Sinitsyn}\ and\ \citenamefont {Li}(2016)}]{DTCM}%
  \BibitemOpen
  \bibfield  {author} {\bibinfo {author} {\bibfnamefont {Nikolai~A.}\
  \bibnamefont {Sinitsyn}}\ and\ \bibinfo {author} {\bibfnamefont {Fuxiang}\
  \bibnamefont {Li}},\ }\bibfield  {title} {\enquote {\bibinfo {title}
  {Solvable multistate model of {Landau-Zener} transitions in cavity {QED}},}\
  }\href {\doibase 10.1103/PhysRevA.93.063859} {\bibfield  {journal} {\bibinfo
  {journal} {Phys. Rev. A}\ }\textbf {\bibinfo {volume} {93}},\ \bibinfo
  {pages} {063859} (\bibinfo {year} {2016})}\BibitemShut {NoStop}%
\bibitem [{\citenamefont {Sun}\ \emph {et~al.}(2019)\citenamefont {Sun},
  \citenamefont {Chernyak}, \citenamefont {Piryatinski},\ and\ \citenamefont
  {Sinitsyn}}]{laser-LZ}%
  \BibitemOpen
  \bibfield  {author} {\bibinfo {author} {\bibfnamefont {Chen}\ \bibnamefont
  {Sun}}, \bibinfo {author} {\bibfnamefont {Vladimir~Y.}\ \bibnamefont
  {Chernyak}}, \bibinfo {author} {\bibfnamefont {Andrei}\ \bibnamefont
  {Piryatinski}}, \ and\ \bibinfo {author} {\bibfnamefont {Nikolai~A.}\
  \bibnamefont {Sinitsyn}},\ }\bibfield  {title} {\enquote {\bibinfo {title}
  {Cooperative light emission in the presence of strong inhomogeneous
  broadening},}\ }\href {\doibase 10.1103/PhysRevLett.123.123605} {\bibfield
  {journal} {\bibinfo  {journal} {Phys. Rev. Lett.}\ }\textbf {\bibinfo
  {volume} {123}},\ \bibinfo {pages} {123605} (\bibinfo {year}
  {2019})}\BibitemShut {NoStop}%
\bibitem [{\citenamefont {Radzihovsky}(2004)}]{phase-transition-1}%
  \BibitemOpen
  \bibfield  {author} {\bibinfo {author} {\bibfnamefont {Park J. Weichman
  P.~B.}\ \bibnamefont {Radzihovsky}, \bibfnamefont {L.}},\ }\bibfield  {title}
  {\enquote {\bibinfo {title} {Superfluid transitions in bosonic atom-molecule
  mixtures near a {Feshbach} resonance},}\ }\href@noop {} {\bibfield  {journal}
  {\bibinfo  {journal} {Phys. Rev. Lett.}\ }\textbf {\bibinfo {volume} {92}},\
  \bibinfo {pages} {160402} (\bibinfo {year} {2004})}\BibitemShut {NoStop}%
\bibitem [{\citenamefont {Romans}\ \emph {et~al.}(2004)\citenamefont {Romans},
  \citenamefont {Duine}, \citenamefont {Sachdev},\ and\ \citenamefont
  {Stoof}}]{phase-transition-2}%
  \BibitemOpen
  \bibfield  {author} {\bibinfo {author} {\bibfnamefont {M.~W.~J.}\
  \bibnamefont {Romans}}, \bibinfo {author} {\bibfnamefont {R.~A.}\
  \bibnamefont {Duine}}, \bibinfo {author} {\bibfnamefont {Subir}\ \bibnamefont
  {Sachdev}}, \ and\ \bibinfo {author} {\bibfnamefont {H.~T.~C.}\ \bibnamefont
  {Stoof}},\ }\bibfield  {title} {\enquote {\bibinfo {title} {Quantum phase
  transition in an atomic {Bose} gas with a {Feshbach} resonance},}\ }\href
  {\doibase 10.1103/PhysRevLett.93.020405} {\bibfield  {journal} {\bibinfo
  {journal} {Phys. Rev. Lett.}\ }\textbf {\bibinfo {volume} {93}},\ \bibinfo
  {pages} {020405} (\bibinfo {year} {2004})}\BibitemShut {NoStop}%
\bibitem [{\citenamefont {Santos}\ \emph {et~al.}(2006)\citenamefont {Santos},
  \citenamefont {Tonel}, \citenamefont {Foerster},\ and\ \citenamefont
  {Links}}]{angela1}%
  \BibitemOpen
  \bibfield  {author} {\bibinfo {author} {\bibfnamefont {G.}~\bibnamefont
  {Santos}}, \bibinfo {author} {\bibfnamefont {A.}~\bibnamefont {Tonel}},
  \bibinfo {author} {\bibfnamefont {A.}~\bibnamefont {Foerster}}, \ and\
  \bibinfo {author} {\bibfnamefont {J.}~\bibnamefont {Links}},\ }\bibfield
  {title} {\enquote {\bibinfo {title} {Classical and quantum dynamics of a
  model for atomic-molecular bose-einstein condensates},}\ }\href {\doibase
  10.1103/PhysRevA.73.023609} {\bibfield  {journal} {\bibinfo  {journal} {Phys.
  Rev. A}\ }\textbf {\bibinfo {volume} {73}},\ \bibinfo {pages} {023609}
  (\bibinfo {year} {2006})}\BibitemShut {NoStop}%
\bibitem [{\citenamefont {Duncan}\ \emph {et~al.}(2007)\citenamefont {Duncan},
  \citenamefont {Foerster}, \citenamefont {Links}, \citenamefont {Mattei},
  \citenamefont {Oelkers},\ and\ \citenamefont {Tonel}}]{angela2}%
  \BibitemOpen
  \bibfield  {author} {\bibinfo {author} {\bibfnamefont {Melissa}\ \bibnamefont
  {Duncan}}, \bibinfo {author} {\bibfnamefont {Angela}\ \bibnamefont
  {Foerster}}, \bibinfo {author} {\bibfnamefont {Jon}\ \bibnamefont {Links}},
  \bibinfo {author} {\bibfnamefont {Eduardo}\ \bibnamefont {Mattei}}, \bibinfo
  {author} {\bibfnamefont {Norman}\ \bibnamefont {Oelkers}}, \ and\ \bibinfo
  {author} {\bibfnamefont {Arlei~Prestes}\ \bibnamefont {Tonel}},\ }\bibfield
  {title} {\enquote {\bibinfo {title} {Emergent quantum phases in a
  heteronuclear molecular bose–einstein condensate model},}\ }\href {\doibase
  https://doi.org/10.1016/j.nuclphysb.2006.12.015} {\bibfield  {journal}
  {\bibinfo  {journal} {Nuclear Physics B}\ }\textbf {\bibinfo {volume}
  {767}},\ \bibinfo {pages} {227--249} (\bibinfo {year} {2007})}\BibitemShut
  {NoStop}%
\bibitem [{\citenamefont {Santos}\ \emph {et~al.}(2010)\citenamefont {Santos},
  \citenamefont {Foerster}, \citenamefont {Links}, \citenamefont {Mattei},\
  and\ \citenamefont {Dahmen}}]{angela3}%
  \BibitemOpen
  \bibfield  {author} {\bibinfo {author} {\bibfnamefont {G.}~\bibnamefont
  {Santos}}, \bibinfo {author} {\bibfnamefont {A.}~\bibnamefont {Foerster}},
  \bibinfo {author} {\bibfnamefont {J.}~\bibnamefont {Links}}, \bibinfo
  {author} {\bibfnamefont {E.}~\bibnamefont {Mattei}}, \ and\ \bibinfo {author}
  {\bibfnamefont {S.~R.}\ \bibnamefont {Dahmen}},\ }\bibfield  {title}
  {\enquote {\bibinfo {title} {Quantum phase transitions in an interacting
  atom-molecule boson model},}\ }\href {\doibase 10.1103/PhysRevA.81.063621}
  {\bibfield  {journal} {\bibinfo  {journal} {Phys. Rev. A}\ }\textbf {\bibinfo
  {volume} {81}},\ \bibinfo {pages} {063621} (\bibinfo {year}
  {2010})}\BibitemShut {NoStop}%
\bibitem [{\citenamefont {Sinitsyn}(2013)}]{sinitsyn-chain}%
  \BibitemOpen
  \bibfield  {author} {\bibinfo {author} {\bibfnamefont {N.~A.}\ \bibnamefont
  {Sinitsyn}},\ }\bibfield  {title} {\enquote {\bibinfo {title} {{Landau-Zener}
  transitions in chains},}\ }\href {\doibase 10.1103/PhysRevA.87.032701}
  {\bibfield  {journal} {\bibinfo  {journal} {Phys. Rev. A}\ }\textbf {\bibinfo
  {volume} {87}},\ \bibinfo {pages} {032701} (\bibinfo {year}
  {2013})}\BibitemShut {NoStop}%
\bibitem [{\citenamefont {Rangelov}\ \emph {et~al.}(2005)\citenamefont
  {Rangelov}, \citenamefont {Piilo},\ and\ \citenamefont
  {Vitanov}}]{vitanov-parallel}%
  \BibitemOpen
  \bibfield  {author} {\bibinfo {author} {\bibfnamefont {A.~A.}\ \bibnamefont
  {Rangelov}}, \bibinfo {author} {\bibfnamefont {J.}~\bibnamefont {Piilo}}, \
  and\ \bibinfo {author} {\bibfnamefont {N.~V.}\ \bibnamefont {Vitanov}},\
  }\bibfield  {title} {\enquote {\bibinfo {title} {Counterintuitive transitions
  between crossing energy levels},}\ }\href {\doibase
  10.1103/PhysRevA.72.053404} {\bibfield  {journal} {\bibinfo  {journal} {Phys.
  Rev. A}\ }\textbf {\bibinfo {volume} {72}},\ \bibinfo {pages} {053404}
  (\bibinfo {year} {2005})}\BibitemShut {NoStop}%
\end{thebibliography}%


\begin{thebibliography}{7}%
\makeatletter
\providecommand \@ifxundefined [1]{%
 \@ifx{#1\undefined}
}%
\providecommand \@ifnum [1]{%
 \ifnum #1\expandafter \@firstoftwo
 \else \expandafter \@secondoftwo
 \fi
}%
\providecommand \@ifx [1]{%
 \ifx #1\expandafter \@firstoftwo
 \else \expandafter \@secondoftwo
 \fi
}%
\providecommand \natexlab [1]{#1}%
\providecommand \enquote  [1]{``#1''}%
\providecommand \bibnamefont  [1]{#1}%
\providecommand \bibfnamefont [1]{#1}%
\providecommand \citenamefont [1]{#1}%
\providecommand \href@noop [0]{\@secondoftwo}%
\providecommand \href [0]{\begingroup \@sanitize@url \@href}%
\providecommand \@href[1]{\@@startlink{#1}\@@href}%
\providecommand \@@href[1]{\endgroup#1\@@endlink}%
\providecommand \@sanitize@url [0]{\catcode `\\12\catcode `\$12\catcode
  `\&12\catcode `\#12\catcode `\^12\catcode `\_12\catcode `\%12\relax}%
\providecommand \@@startlink[1]{}%
\providecommand \@@endlink[0]{}%
\providecommand \url  [0]{\begingroup\@sanitize@url \@url }%
\providecommand \@url [1]{\endgroup\@href {#1}{\urlprefix }}%
\providecommand \urlprefix  [0]{URL }%
\providecommand \Eprint [0]{\href }%
\providecommand \doibase [0]{http://dx.doi.org/}%
\providecommand \selectlanguage [0]{\@gobble}%
\providecommand \bibinfo  [0]{\@secondoftwo}%
\providecommand \bibfield  [0]{\@secondoftwo}%
\providecommand \translation [1]{[#1]}%
\providecommand \BibitemOpen [0]{}%
\providecommand \bibitemStop [0]{}%
\providecommand \bibitemNoStop [0]{.\EOS\space}%
\providecommand \EOS [0]{\spacefactor3000\relax}%
\providecommand \BibitemShut  [1]{\csname bibitem#1\endcsname}%
\let\auto@bib@innerbib\@empty
\bibitem [{\citenamefont {Yurovsky}\ \emph {et~al.}(2002)\citenamefont
  {Yurovsky}, \citenamefont {Ben-Reuven},\ and\ \citenamefont
  {Julienne}}]{yurovsky}%
  \BibitemOpen
  \bibfield  {author} {\bibinfo {author} {\bibfnamefont {V.~A.}\ \bibnamefont
  {Yurovsky}}, \bibinfo {author} {\bibfnamefont {A.}~\bibnamefont
  {Ben-Reuven}}, \ and\ \bibinfo {author} {\bibfnamefont {P.~S.}\ \bibnamefont
  {Julienne}},\ }\bibfield  {title} {\enquote {\bibinfo {title} {Quantum
  effects on curve crossing in a {Bose-Einstein} condensate},}\ }\href
  {\doibase 10.1103/PhysRevA.65.043607} {\bibfield  {journal} {\bibinfo
  {journal} {Phys. Rev. A}\ }\textbf {\bibinfo {volume} {65}},\ \bibinfo
  {pages} {043607} (\bibinfo {year} {2002})}\BibitemShut {NoStop}%
\bibitem [{\citenamefont {Kayali}\ and\ \citenamefont
  {Sinitsyn}(2003)}]{kayali}%
  \BibitemOpen
  \bibfield  {author} {\bibinfo {author} {\bibfnamefont {M.~A.}\ \bibnamefont
  {Kayali}}\ and\ \bibinfo {author} {\bibfnamefont {N.~A.}\ \bibnamefont
  {Sinitsyn}},\ }\bibfield  {title} {\enquote {\bibinfo {title} {Formation of a
  two-component {Bose} condensate during the chemical-potential curve
  crossing},}\ }\href {\doibase 10.1103/PhysRevA.67.045603} {\bibfield
  {journal} {\bibinfo  {journal} {Phys. Rev. A}\ }\textbf {\bibinfo {volume}
  {67}},\ \bibinfo {pages} {045603} (\bibinfo {year} {2003})}\BibitemShut
  {NoStop}%
\bibitem [{\citenamefont {Sinitsyn}(2013)}]{sinitsyn-chain}%
  \BibitemOpen
  \bibfield  {author} {\bibinfo {author} {\bibfnamefont {N.~A.}\ \bibnamefont
  {Sinitsyn}},\ }\bibfield  {title} {\enquote {\bibinfo {title} {{Landau-Zener}
  transitions in chains},}\ }\href {\doibase 10.1103/PhysRevA.87.032701}
  {\bibfield  {journal} {\bibinfo  {journal} {Phys. Rev. A}\ }\textbf {\bibinfo
  {volume} {87}},\ \bibinfo {pages} {032701} (\bibinfo {year}
  {2013})}\BibitemShut {NoStop}%
\bibitem [{\citenamefont {Sinitsyn}\ and\ \citenamefont {Li}(2016)}]{DTCM}%
  \BibitemOpen
  \bibfield  {author} {\bibinfo {author} {\bibfnamefont {Nikolai~A.}\
  \bibnamefont {Sinitsyn}}\ and\ \bibinfo {author} {\bibfnamefont {Fuxiang}\
  \bibnamefont {Li}},\ }\bibfield  {title} {\enquote {\bibinfo {title}
  {Solvable multistate model of {Landau-Zener} transitions in cavity {QED}},}\
  }\href {\doibase 10.1103/PhysRevA.93.063859} {\bibfield  {journal} {\bibinfo
  {journal} {Phys. Rev. A}\ }\textbf {\bibinfo {volume} {93}},\ \bibinfo
  {pages} {063859} (\bibinfo {year} {2016})}\BibitemShut {NoStop}%
\bibitem [{\citenamefont {Sun}\ and\ \citenamefont {Sinitsyn}(2016)}]{DTCM1}%
  \BibitemOpen
  \bibfield  {author} {\bibinfo {author} {\bibfnamefont {Chen}\ \bibnamefont
  {Sun}}\ and\ \bibinfo {author} {\bibfnamefont {Nikolai~A.}\ \bibnamefont
  {Sinitsyn}},\ }\bibfield  {title} {\enquote {\bibinfo {title} {{Landau-Zener}
  extension of the {Tavis-Cummings} model: Structure of the solution},}\ }\href
  {\doibase 10.1103/PhysRevA.94.033808} {\bibfield  {journal} {\bibinfo
  {journal} {Phys. Rev. A}\ }\textbf {\bibinfo {volume} {94}},\ \bibinfo
  {pages} {033808} (\bibinfo {year} {2016})}\BibitemShut {NoStop}%
\bibitem [{\citenamefont {Sinitsyn}\ \emph {et~al.}(2018)\citenamefont
  {Sinitsyn}, \citenamefont {Yuzbashyan}, \citenamefont {Chernyak},
  \citenamefont {Patra},\ and\ \citenamefont {Sun}}]{commute}%
  \BibitemOpen
  \bibfield  {author} {\bibinfo {author} {\bibfnamefont {Nikolai~A.}\
  \bibnamefont {Sinitsyn}}, \bibinfo {author} {\bibfnamefont {Emil~A.}\
  \bibnamefont {Yuzbashyan}}, \bibinfo {author} {\bibfnamefont {Vladimir~Y.}\
  \bibnamefont {Chernyak}}, \bibinfo {author} {\bibfnamefont {Aniket}\
  \bibnamefont {Patra}}, \ and\ \bibinfo {author} {\bibfnamefont {Chen}\
  \bibnamefont {Sun}},\ }\bibfield  {title} {\enquote {\bibinfo {title}
  {Integrable time-dependent quantum {Hamiltonians}},}\ }\href {\doibase
  10.1103/PhysRevLett.120.190402} {\bibfield  {journal} {\bibinfo  {journal}
  {Phys. Rev. Lett.}\ }\textbf {\bibinfo {volume} {120}},\ \bibinfo {pages}
  {190402} (\bibinfo {year} {2018})}\BibitemShut {NoStop}%
\bibitem [{\citenamefont {Rangelov}\ \emph {et~al.}(2005)\citenamefont
  {Rangelov}, \citenamefont {Piilo},\ and\ \citenamefont
  {Vitanov}}]{vitanov-parallel}%
  \BibitemOpen
  \bibfield  {author} {\bibinfo {author} {\bibfnamefont {A.~A.}\ \bibnamefont
  {Rangelov}}, \bibinfo {author} {\bibfnamefont {J.}~\bibnamefont {Piilo}}, \
  and\ \bibinfo {author} {\bibfnamefont {N.~V.}\ \bibnamefont {Vitanov}},\
  }\bibfield  {title} {\enquote {\bibinfo {title} {Counterintuitive transitions
  between crossing energy levels},}\ }\href {\doibase
  10.1103/PhysRevA.72.053404} {\bibfield  {journal} {\bibinfo  {journal} {Phys.
  Rev. A}\ }\textbf {\bibinfo {volume} {72}},\ \bibinfo {pages} {053404}
  (\bibinfo {year} {2005})}\BibitemShut {NoStop}%
\end{thebibliography}%




\end{document}


\title{Supplementary material for ``Coherent reaction between molecular and atomic Bose-Einstein condensates: integrable model"}
\author{Rajesh K. Malla}
\affiliation{Theoretical Division, and the Center for Nonlinear Studies, Los Alamos National Laboratory, Los Alamos, New Mexico 87545, USA}

\author{Vladimir Y. Chernyak}
\affiliation{Department of Chemistry, Wayne State University, 5101 Cass Ave, Detroit, Michigan 48202, USA}
\affiliation{Department of Mathematics, Wayne State University, 656 W. Kirby, Detroit, Michigan 48202, USA}
\author{Chen Sun}
\affiliation{Hunan University, China}
\author{Nikolai A. Sinitsyn}
\affiliation{Theoretical Division, Los Alamos National Laboratory, Los Alamos, New Mexico 87545, USA}

\maketitle
\section{Molecular dissociation and production in the strongly nonadiabatic regime}
 
 In the main text, we did not discuss the limit of fast transitions, such that the initial state remains practically intact but a considerable number of particles of a new type is generated.
 This regime in our model reduces to simple models that have been studied previously, so we briefly review the earlier work here for completeness. 

For the reverse sweep, the number of molecules does not change much in the strongly nonadiabatic regime, so we can replace the field $\hat{\Psi}$ in the interaction term in (1) of the main text by a constant c-number (Ref.~\cite{yurovsky,kayali} in this supplementary).  
This transforms the original Hamiltonian  to  a quadratic Hamiltonian for only the atomic fields $\hat{a}$ and $\hat{b}$, without interactions between the different reaction channels. Ref.~\cite{yurovsky}  assumed that only one atomic mode is interacting with the molecular condensate, i.e., the effective Hamiltonian was reduced to
\be
H=\beta t \hat{a}^{\dagger} \hat{a} + \gamma  \hat{a}^2+\gamma^* (\hat{a}^{\dagger})^2,
\label{Y-h1}
\ee
where $\gamma=g\la \hat{\Psi} \ra$. Reference~\cite{kayali} started from a more general model of a broad Feschbach resonance. It, however, included the main text's Eq.~(1) as a special case. In the nonadiabatic regime the effective Hamiltonian was found to be
\be
H=\beta t \hat{a}^{\dagger} \hat{a} + \gamma  \hat{a}\hat{b}+\gamma^* \hat{a}^{\dagger}\hat{b}^{\dagger},
\label{K-h1}
\ee
where $\hat{a}$ and $\hat{b}$ are generally different atomic modes. 

We refer for the analysis of  (\ref{Y-h1}) and (\ref{K-h1}) to the original publications \cite{yurovsky,kayali}, which found exponentially quickly growing reaction efficiency as function of the inverse sweep rate $\beta$ and squeezing in the atomic state after the reverse process.  The forward process, from atoms to molecules, was also studied in the same strongly nonadiabatic regime by treating the molecular field $\hat{\Psi}$ as a quantum operator and the atomic fields as c-numbers in \cite{sinitsyn-chain}, which found that the molecular state appears as a bosonic coherent state.

\section{Scattering Phases in Degenerate Tavis-Cummings model}

The degenerate Tavis-Cummings model, Eq.~(8) in the main text, is found as the limit of the standard driven Tavis-Cummings model that describes interaction of $N$ spins-1/2 with a bosonic mode \cite{DTCM,DTCM1}:
\begin{equation}
H=-\beta t   \hat{a}^{\dagger}\hat{a} +\sum_{k=1}^N \varepsilon_k \sigma_k^z + \sum_{k=1}^N g(\hat{a}^{\dagger} \sigma_k^- +\hat{a} \sigma_k^+),
\label{dtcm1}
\end{equation}
where $\hat{a}$ is the boson annihilation operator; $\sigma_k$ are the Pauli operators of $N$ spins;  $\varepsilon_k$ and $g$ is the spin-boson coupling.
\begin{figure}
\includegraphics[scale=0.25]{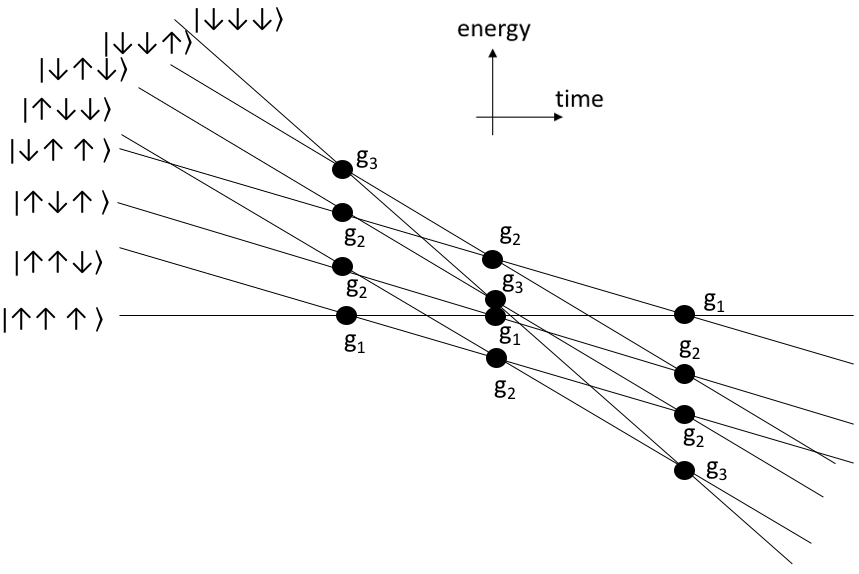}
\caption{Diabatic energy levels, i.e., the time-dependent diagonal elements of the Tavis-Cummings Hamiltonian (\ref{dtcm1}) in the basis of spins along $z$-axis, for $N=3$ spins. The couplings are $g_n=g\sqrt{n+Q}$, where $Q$ is an independent parameter of the model.
}
\label{TC-levels}
\end{figure}

Let $Q$ be the number of bosons in the fully polarized-up spin state. Then, the matrix form of this Hamiltonian has linearly time-dependent diagonal elements (diabatic levels) that represent the states, which are coupled by fixed couplings
$$
g_n=g\sqrt{n+Q},
$$
where $n$ is the number of the additional bosons that are emitted to reach the given state from the fully polarized state. Then, $g_n$ is the coupling of this state with the level that has one more spin up.
In Fig.~\ref{TC-levels}, we show such diabatic levels and the corresponding couplings between them for $N=3$. We refer to \cite{DTCM} for details about construction of the diabatic level diagram.

The model (\ref{dtcm1}) is exactly solvable, and the amplitude of each final state is given by the sum of amplitudes of the semiclassical trajectories that connect the initial diabatic state as $t\rightarrow -\infty$ to the final state as $t\rightarrow +\infty$ \cite{commute,DTCM}. The amplitude of each trajectory is then found as the product of the Landau-Zener transition amplitudes applied to each level intersection encountered along the trajectory \cite{commute}. In this sense, both the transition probabilities and scattering phases in this model are known, although only the probabilities have been studied in detail in the previous research \cite{DTCM1}.

Another consequence of the integrability is that the scattering amplitudes do not depend on the parameters $\varepsilon_k$, apart from their relative order. 
This property was used in \cite{DTCM} to derive the transition probabilities in the degenerate version of the Tavis-Cummings model.

Namely, for all $\varepsilon_k=0$, only the fully symmetrized superpositions of the diabatic spin states are coupled to each other, so the model (\ref{dtcm1}) splits into independent sectors of the degenerate model in Eq~(8) of the main text. Hence, the solution of the degenerate version can be obtained from (\ref{dtcm1}) by setting $\varepsilon_k=\tau\varepsilon_k$, and taking the limit $\tau\rightarrow 0$. 
However, this limit has to be taken with extra care because for any finite $\varepsilon_k$, there are parallel levels in the diabatic level diagram (see Fig.~\ref{TC-levels}). Between them, the transitions happen during the time that diverges in the degenerate limit \cite{vitanov-parallel}, so in the spin sectors that contain such parallel diabatic levels, additional investigation of this slow ``equilibration" is needed to connect the amplitudes of the states in the degenerate and nondegenerate models. 

Nevertheless, the model (\ref{dtcm1}) always has two diabatic levels that do not belong to a band of more than one parallel levels. They are the levels with the lowest and the highest slopes. For their amplitudes, the limit $\tau \rightarrow 0$ in (\ref{dtcm1}) is taken trivially. This leads to the property that the amplitudes of the transitions between them in the degenerate and nondegenerate version of the Tavis-Cummings model coincide. 

Thus,  we can obtain the scattering amplitude of the ``corner-to-corner transition", i.e., of the process that corresponds to the transition from all spins initially polarized up to the state with all spins polarized down. In figure~\ref{dtcm1}  there is a single semiclassical trajectory that connects these two states, and this property is satisfied for any $N$ \cite{DTCM}. On the way, the system encounters $N$ level crossings with the couplings in the range $g_1,\ldots,g_{N}$.
Hence, the probability of such a transition is the product of the Landau-Zener turning probabilities: 
$$
P_{|\uparrow \ldots \uparrow \rangle \rightarrow|\downarrow \ldots \downarrow \rangle} =\prod_{k=1}^{N} (1-e^{-2\pi g_k^2/\beta}).
$$
The corresponding scattering phase is the sum of the Landau-Zener phases:
\begin{equation}
\phi_N=\sum_{k=1}^{N} \varphi_k.
\label{pcc}
\end{equation}
For the two-state Landau-Zener model, the phase $\varphi_k$ for switching between  two diabatic states is known analytically \cite{yurovsky}. Up to an arbitrary $2\pi n$ addition, it is given by 
\begin{equation}
\varphi_k=3\pi/4 - arg\Gamma[ig_k^2/\beta].
\label{pcc1}
\end{equation}

Using the map from the Tavis-Cummings model to the model of the BEC reaction, we find that this phase also describes the scattering phase for the amplitude of the complete molecular dissociation into the lowest energy atomic mode. The case of a single such atomic mode ($\hat{a} \equiv \hat{b}$) without atoms in the initial state corresponds to $Q=1$. Thus, we recover Eq.~(20) in the main text.

\section{The average number of molecules at finite Q}
In the main text, Figs.~1(c,d), and the expression  for the average number of the produced molecules (10) were shown for finite values of the conserved quantity $Q$. Previously, this situation has not been considered for the Tavis-Cummings model in \cite{DTCM1}, apart from  the most general exact formulas. The reason was that, for the fermion-boson reactions, $Q\ne 0$ corresponded to a finite molecular population before the forward sweep, which was experimentally irrelevant. However, for the Hamiltonian~(6) in the main text, $Q\ne 0$ corresponds to a physically expected case with a small  difference between initial populations of $a/b$ atomic modes. In addition, the case with $Q=1$ corresponds to $\hat{a}\equiv \hat{b}$ situation, which 
is relevant to the initial condition with a single atomic BEC. Hence, here we provide exact formulas and explore the large-$N$ limit at $Q\ne 0$ for the forward process.

The processes in the Tavis-Cummings model and our model~(6) are switched, namely, the forward process in the Tavis-Cummings model is the reverse process in our case and vice versa. We define, $P_{n0}$ is the probability to produce $n$ molecules from $0$ molecules, and $P_{nN}$ is the probability to produce $N-n$ atomic pairs from initially prepared molecular condensate with $N$ molecules.  At the end of the forward and reverse sweeps they  are given, respectively, by \cite{DTCM1} 
\begin{eqnarray}
P_{n0}=\begin{pmatrix}
N\\N-n
\end{pmatrix}_{x} x^{(N+Q-n)n}(x^{Q+N-n+1},x)_{n},
\label{prob-forward}\\
P_{nN}=\begin{pmatrix}
N\\n
\end{pmatrix}_{x} x^{(Q+1)(N-n)}(x^{Q+1},x)_{n},
\label{prob-reverse}
\end{eqnarray}
where we assumed no molecules initially for the forward sweep,  and no atoms in one of the atomic modes for the reverse sweep, and where 
$$
\begin{pmatrix}
m\\k
\end{pmatrix}_{x}=\frac{(x,x)_m}{(x,x)_k(x,x)_{m-k}};
$$
 \begin{equation}
    (a,x)_m\equiv \prod_{k=0}^{m-1}(1-ax^k) = (1-a)(1-ax)\cdots(1-ax^{m-1}).
    \label{qPotch-def}
\end{equation}
is the q-Pochhammer symbol. 

The expression~(11) in the main text follows from (\ref{prob-reverse})  when we substitute  $Q=0$.



\begin{figure}
\includegraphics[scale=0.25]{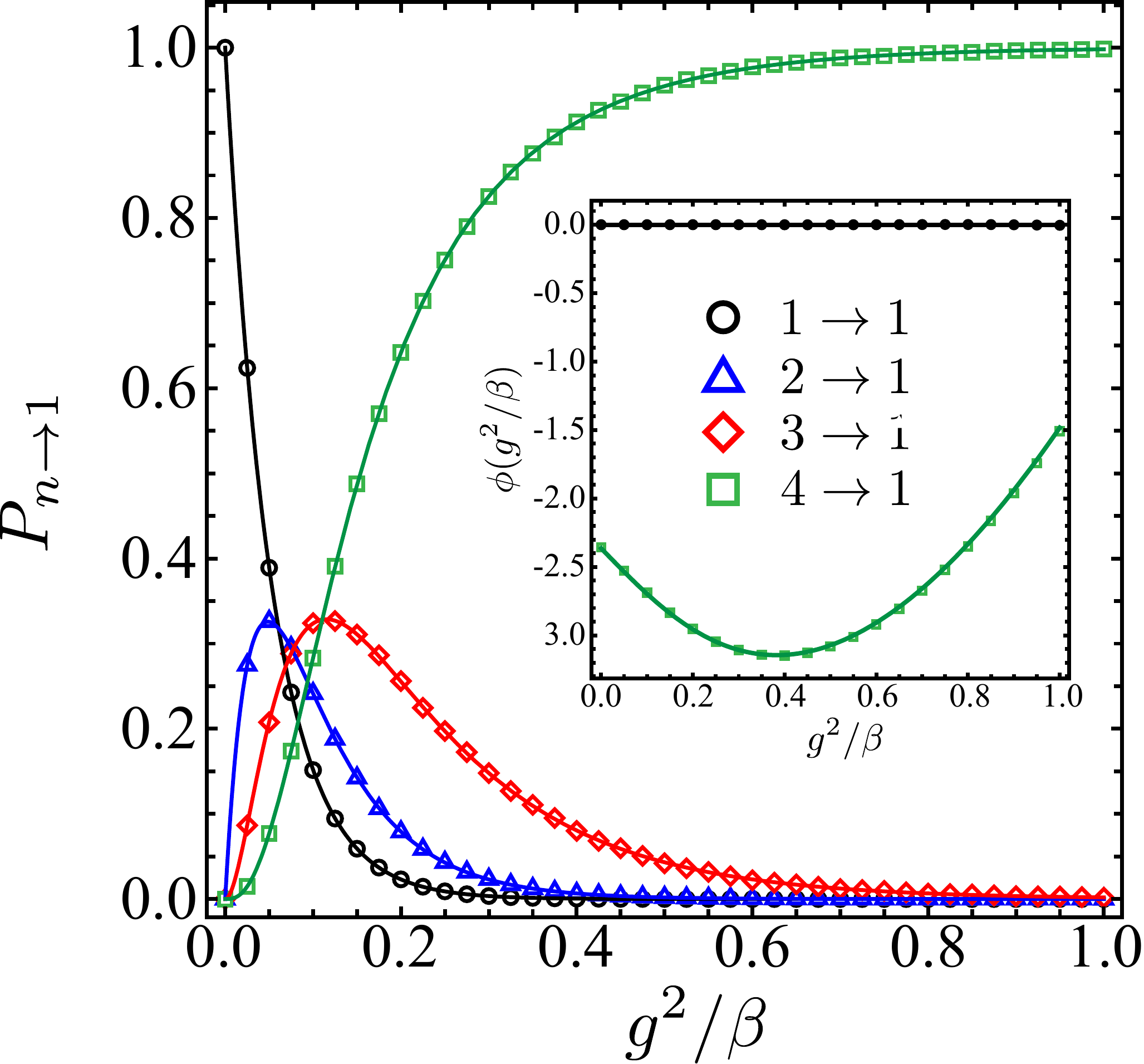}
\caption{The transition probabilities $P_{n\rightarrow 1}$, where $n$ is the index of the diabatic state (not the number of molecules) in (\ref{mat-4}),
as functions of $g^2/|\beta|$ when $\beta>0$. The discrete points represent numerically calculated values. The solid curves represent the analytical predictions of our solution for this sector.  The inset shows the test for the phase (\ref{pcc}),~(\ref{pcc1}) formulas for the scattering matrix element $S_{4\rightarrow 1}$.
}
\label{check-prob}
\end{figure}
\begin{figure}
\includegraphics[scale=0.25]{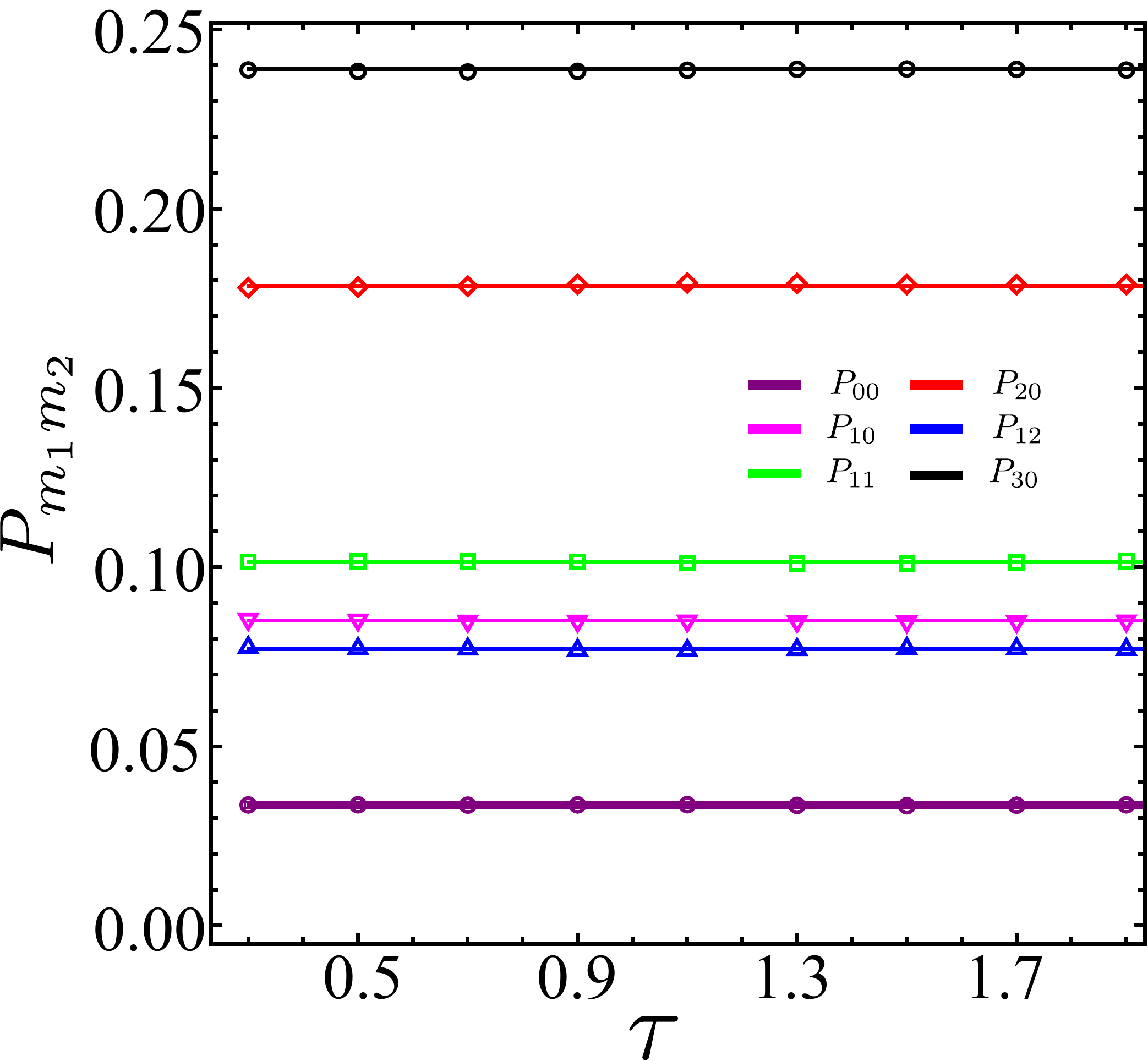}
\caption{{\it Test of Integrability}. Numerically obtained $\tau$-dependence of the joint probabilities $P_{m_1,m_2}$ (plot markers) values for the Hamiltonian~(\ref{mat-10}). Solid lines are the corresponding theoretical predictions in Eq.~(15) of the main text. Here, the parameters of the Hamiltonian~(\ref{mat-10}) are: $g^2/|\beta|=0.3$, $ \varepsilon_{1}=0.5$, and $\varepsilon_{2}=1$. Numerical evolution is performed  in a time frame $[-1000,1000]$ with a time step of $\Delta t=0.0001$. 
This figure confirms the independence of the transition probabilities of the energy rescaling by $\tau$, which is the direct nontrivial consequence of the integrability.}
\label{check-tau}
\end{figure}

\begin{figure}
\includegraphics[scale=0.25]{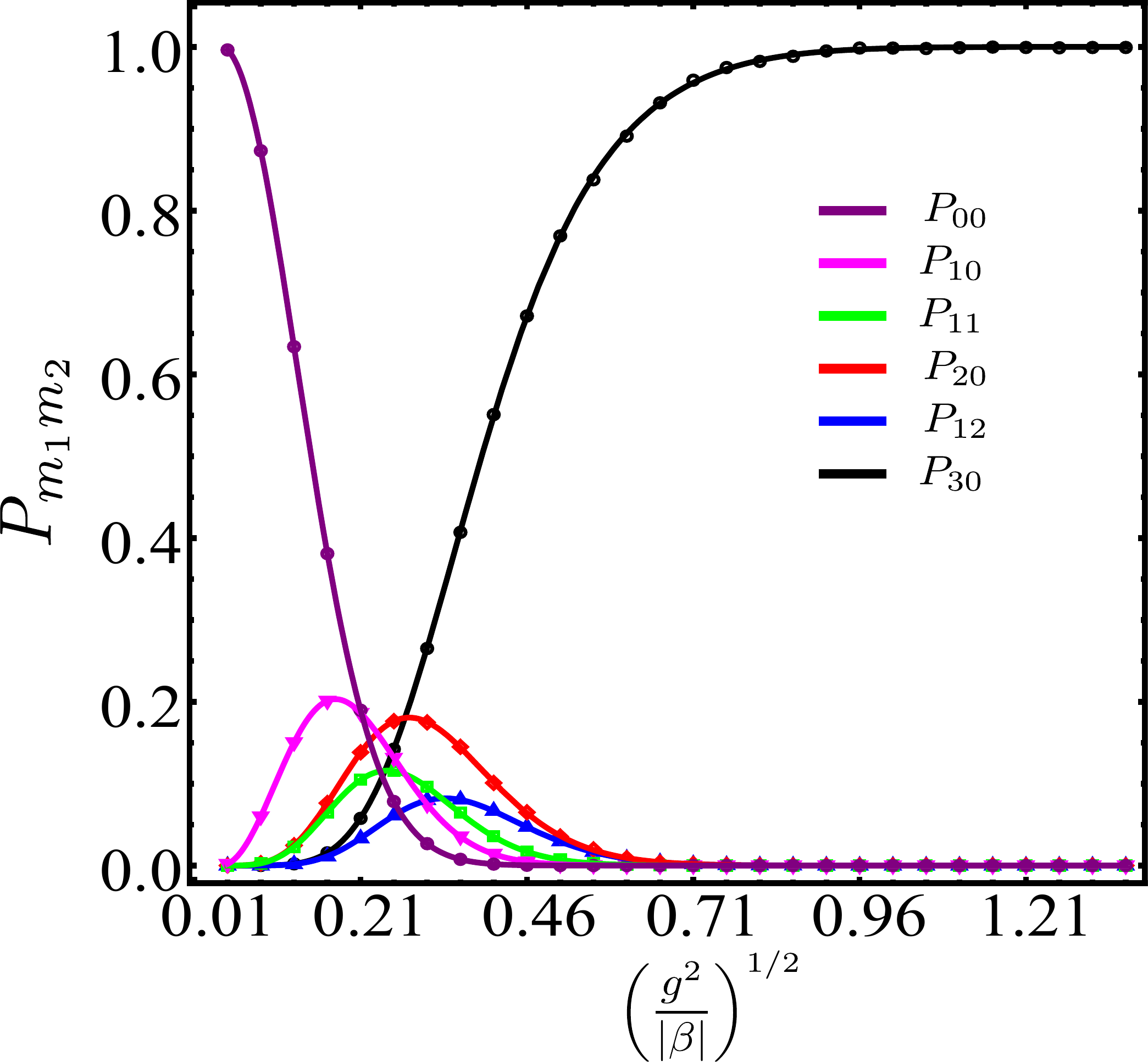}
\caption{{\it Test of Eq.~(15) in the main text.} The joint probabilities $P_{m1m2}$ were obtained by solving the nonstationary Schr\"odinger equation numerically (plot markers) for the Hamiltonian~(\ref{mat-10}).  Solid lines are the corresponding predictions of Eq.~(15) from the main text.
The parameters in (\ref{mat-10}) are:  $\tau=0.7$, $\varepsilon_{1}=0.5$, and $\varepsilon_{2}=1$. 
}
\label{check-jointp}
\end{figure}
{\it Large-N limit for the forward process}.
The probability (\ref{prob-forward}) to find $n$ molecules after the forward sweep satisfies a recursive formula.  Let us introduce a new variable $m=N-n$, and express the recursive formula in the form
\be
{\cal P}(m+1)=\frac{x^Q\left(x^{2m+1}-x^{N+m+1} \right)}{\left(1-x^{m+1} \right)\left(1-x^{m+Q+1}\right)}{\cal P}(m).
\label{recursive}
\ee
We treat $m$ as a continuous variable and obtain the differential equation for the probability density 
\begin{multline}
\frac{1}{{\cal P}} \frac{d {\cal P}}{dm}\Big|_{m+1/2}=2\frac{{\cal P}_{m+1}-{\cal P}_{m}}{{\cal P}_{m+1}+{\cal P}_{m}}=-2\\
+2\left( \frac{2x^{Q}\left(x^{2m+1}-x^{N+m+1} \right)}{x^{Q}\left(x^{2m+1}-x^{N+m+1} \right)+\left(1-x^{m+1} \right)\left(1-x^{m+Q+1}\right)}\right).
\label{diff2}
\end{multline}
The differential equation (\ref{diff2}) has an exact solution. We find the average number of molecules in the large $N$ limit by calculating the maximum of the solution, which requires solving the following equation
\be
1-2x^{m+1}+x^{2m+2}-x^{2m+Q+1}+x^{N+m+Q+1}=0.
\label{maximum}
\ee
Since $m$ is large we can eliminate the terms with $x^{2m}$, and reproduce the expression for the average number of molecules:
\be
\langle n\rangle=N+\frac{\log{\left(2-x^{N+Q}\right)}}{\log{x}},
\ee
which is Eq.~(10) in the main text. 

Analogous study of the large-$N$ limit for the reverse process predicts the maximum of the distribution (4) at 
\be
\langle n \rangle=N-\frac{\log(1-x^{(Q+1)})}{\log x}.
\ee
The average number of atomic pairs formed is $\langle m\rangle=N-\langle n \rangle$, so the expression for the normalized number is 
\be
\frac{\langle m \rangle}{N}=\frac{\log (1-x^{Q+1})}{\log x}.
\label{rev-avg}
\ee
At  $Q=0$ the equation (\ref{rev-avg}) reproduces Eq.~(35)  in [5].

\section{Numerical check for the exact solution}
We tested our analytical predictions, including the integrability of the Hamiltonian~(1)  and the joint probability~(15) in the main text, by solving the Schr{\"o}dinger equations numerically for sufficiently small finite values of $N$ and $Q$.

\subsection{The single channel Hamiltonian~(6) in the main text, and the scattering phase} 
 The $N=3$ case of this model corresponds to a $4\times4$ matrix,
 \be
H_{4}(t)=\begin{pmatrix}
0 & 3g_1 & 0 & 0 \\
3g_1 & -\beta t & 2 g_2 & 0 \\
0 & 2 g_2 & -2\beta t & g_3\\
0 & 0 & g_3 & -3\beta t
\end{pmatrix},
\label{mat-4}
\ee
 whose transition probabilities were studied in Ref.~\cite{DTCM} within the corresponding degenerate Tavis-Cummings model. Numerically we reproduced these results for $N=3$ case in Fig.~\ref{check-prob}. This validates our numerical technique.
 In the inset
 of Fig.~\ref{check-prob}, we also test our prediction for the phase of the scattering matrix element $S_{4\rightarrow 1}$, which is found to be in perfect agreement with Eqs.~(\ref{pcc}),~(\ref{pcc1}).

\subsection{Test of integrability}
To test the integrability of the Hamiltonian~(3) in the main text, we write this Hamiltonian in the matrix form for two reaction channels. Let, $\varepsilon_k$, $k=1,2$, be the atomic mode energies and let us define the coupling parameters. $g_{n}=g\sqrt{n+Q}$. In the sector $N=3$ and $Q=0$ the Hamiltonian~(3) in the main text is a $10\times 10$ matrix:
\begin{widetext}
\begin{equation}
H(t)=\begin{pmatrix}
 3\varepsilon_1 \tau & 0 & 0 & 0 & 3g_{1} & 0 & 0 & 0 & 0 & 0 \\
 0 &  (2\varepsilon_1+\varepsilon_2) \tau & 0 & 0 & g_{1} & 2g_{1} & 0 & 0 & 0 & 0 \\
 0 & 0 & (\varepsilon_1+2\varepsilon_2) \tau & 0 & 0 & 2g_{1} & g_{1} & 0 & 0 & 0 \\
 0 & 0 & 0 & 3\varepsilon_2 \tau & 0 & 0 & 3g_{1} & 0 & 0 & 0 \\
 3g_{1} & g_{1} & 0 & 0 & -\beta t+2\varepsilon_1 \tau & 0 & 0 & 2g_{2} & 0 & 0 \\
0 & 2g_{1} & 2g_{1} & 0 & 0 & -\beta t+(\varepsilon_1+\varepsilon_2) \tau & 0 & g_{2} & g_{2} & 0 \\
0 & 0 & g_{1} & 3g_{1} & 0 & 0 & -\beta t+2\varepsilon_2 \tau  & 0 & 2g_{2} & 0 \\
 0 & 0 & 0 & 0 & 2g_{2} & g_{2} & 0 & -2\beta t+\varepsilon_1 \tau & 0 & g_{3} \\
 0 & 0 & 0 & 0 & 0 & g_{2} & 2g_{2} & 0 & -2\beta t+\varepsilon_2 \tau & g_{3} \\
 0 & 0 & 0 & 0 & 0 & 0 & 0 & g_{3} & g_{3} & -3\beta t 
\end{pmatrix},
    \label{mat-10}
\end{equation}
\end{widetext}


The Schr{\"o}dinger equation was solved using the procedure that was described in the Supplemental Material for Ref.~\cite{DTCM}. 
Figure~\ref{check-tau}  shows that the transition probabilities $P_{10\rightarrow m}$ are independent of  $\tau$ at fixed values of the other parameters. Here, the level $10$ corresponds to the state with three molecules and empty atomic modes. The transition probabilities are independent of energy level splittings $\varepsilon_k$. This confirms  that the Hamiltonian~(3) and the Hamiltonian~(1), in the main text are integrable.  

\subsection{Test of the formula for the joint probabilities $P_{m_1,m_2}$}
Equation~(15) in the main text gives the probabilities to find $m_1$  atomic pairs in the first atomic mode and $m_2$  pairs in the second mode. We test this analytical formula for Hamiltonian (\ref{mat-10}) 
in figure~\ref{check-jointp} 
for the reverse process, starting from only three molecules.  The numerical results are found in perfect agreement with Eq.~(15) in the main text.


\bibliography{ref}